\documentclass[runningheads]{llncs}

\usepackage[
  paperwidth=18.5cm,
  paperheight=30.5cm
]{geometry}

\usepackage{graphicx}
\usepackage{booktabs}
\usepackage{multirow}
\usepackage{url}
\usepackage{array}
\usepackage{amssymb}
\usepackage{tikz}
\usetikzlibrary{arrows.meta,positioning}
\usepackage{amsmath,amssymb,amsfonts}
\usepackage{algorithmic}
\usepackage{graphicx}
\usepackage{textcomp}
\usepackage{amsmath,amssymb,amsfonts}
\usepackage{algorithmic}
\usepackage{graphicx}
\usepackage{textcomp}
\usepackage{xcolor}
\usepackage{tikz}
\usepackage[edges]{forest}
\usepackage{array}
\newcolumntype{C}[1]{>{\centering\arraybackslash}p{#1}}
\newcolumntype{L}[1]{>{\raggedright\arraybackslash}p{#1}}
\usetikzlibrary{arrows.meta,shadows}
\usepackage{enumitem}   % Put this in the preamble
\usepackage{tabularx}
\usepackage{booktabs}   % For beautiful tables
\usepackage{caption}    % For customizing captions
\usepackage{graphicx}
\usetikzlibrary{shapes.geometric, arrows, positioning, chains}
\usepackage{xcolor}
\usepackage{pgfplots}
\usepackage{enumitem}
\def\BibTeX{{\rm B\kern-.05em{\sc i\kern-.025em b}\kern-.08em
    T\kern-.1667em\lower.7ex\hbox{E}\kern-.125emX}}
\usepackage{cite}
\usepackage{multirow}
\usepackage[most]{tcolorbox}
\usepackage[svgnames]{xcolor}
\usepackage[T1]{fontenc}
\setlist[itemize]{itemsep=2pt}
\usepackage{graphicx}
\usepackage{etoolbox}
\makeatletter
\tcbset{
  redditquotebox/.style={
    breakable,
    width=\linewidth,
    top=1pt,
    bottom=1pt,
    boxsep=1pt,
    before skip=6pt,
after skip=6pt,
    boxrule=0.5pt,
        colframe=black,
  }
}

\begin{document}
\title{Students' Perceptions of Peer Grading}

\author{
Uchswas Paul\inst{1} \and
Jash Shah\inst{1} \and
Keira McArthur\inst{1} \and
Aref Babaei\inst{2} \and
Niranjan Rajendran\inst{1} \and
Parvez Rashid\inst{3} \and
Edward Gehringer\inst{1}
}

\authorrunning{Paul et al.}

\institute{
North Carolina State University, Raleigh, NC, USA\\
\email{\{upaul,jshah23,klmcarth,nrajend4,efg\}@ncsu.edu}
\and
Independent Researcher, USA\\
\email{babaei.aref@yahoo.com}
\and
College of Charleston, Charleston, SC, USA\\
\email{rashidp@cofc.edu}
}

\maketitle              % typeset the header of the contribution

\begin{center}
\small\textit{This author version includes additional methodological details that could not be included in the conference paper because of the page limit. The results and conclusions remain unchanged.}
\end{center}

\begin{abstract}
Peer grading is widely used in education, yet it elicits mixed reactions from educators and students. Although many studies have examined students’ views of peer grading, their findings are scattered, and no clear overall picture has emerged. To address this gap, we conducted a mixed-source thematic analysis of literature and student discussions on Reddit. To scale our analysis of the Reddit data, we fine-tuned a Gemini 2.5 text-classification model to classify an initial dataset of 659 posts and 6,607 comments by relevance. Manual review of the items classified as relevant by the model yielded a final dataset of 114 posts and 300 comments. Drawing on evidence from 107 papers and the Reddit dataset, we found that students perceive peer grading as both beneficial and problematic. Positive perceptions included learning and understanding benefits, skill development, engagement, and collaboration, while negative perceptions centered on unreliable grading, unfairness, weak feedback quality, emotional stress, and workload. Reddit discussions also suggested an emerging concern that remains underexplored in the literature: AI use in peer grading may weaken students’ trust in the accuracy and authenticity of the process. We further identified eight mitigation strategies and mapped them to the negative perceptions they help address. Among these, instructor oversight and training played the most central role.

\end{abstract}
\section{Introduction}

Peer assessment is widely recognized as a powerful learning tool, valued for both its scalability and pedagogical benefits. Reviewing peers’ work exposes students to diverse artifacts that clarify expectations and help them calibrate their own performance\cite{paul2025}. A substantial body of research suggests that students often learn more from evaluating others’ work than from reading feedback on their own\cite{Lundstrom2009}, a finding that has been replicated across disciplines and instructional contexts. Studies in writing, engineering, computer science, and the health professions \cite{Double2020} all attest to the learning gains associated with structured peer assessment, including improved metacognitive awareness, deeper engagement with criteria, and enhanced ability to judge quality.

Within this broader landscape, peer grading represents a more contentious subset of peer assessment practices. Instructors often hesitate to delegate grading authority, citing concerns about fairness, reliability, and the loss of professional judgment. The rise of massive open online courses (MOOCs), however, created a context in which peer grading became not just an option but a necessity: with tens of thousands of learners and limited instructional staff, MOOC platforms had no viable alternative for evaluating open-ended work at scale.

Despite the increasing interest in peer assessment, students’ perceptions of peer grading remain poorly synthesized. Existing studies report mixed findings, and prior reviews have focused primarily on technical or procedural aspects, such as anonymity, grading tools, algorithms, and reviewer assignment, rather than students’ experiences\cite{paul2025, paul2026}. To address this gap, our study systematically analyzes students’ perceptions of peer grading using both literature and student discussions on social media. Specifically, we address the following research questions:
\vspace{-2pt}
\begin{itemize}
    \item \textbf{RQ1}: What positive perceptions do students have about peer grading?
    
    \item \textbf{RQ2}: What negative perceptions do students have about peer grading?
    
    \item \textbf{RQ3}: What mitigation strategies are reported in the literature to address peer-grading challenges?
\end{itemize}

\section{Methodology}

We used a mixed‑source thematic design to examine students’ perceptions of peer grading across two complementary sources: peer‑reviewed literature and Reddit discussions. This approach integrates formal research with naturally occurring student discourse. Because peer grading is often discussed within the broader language of peer assessment, we included studies using that terminology when peer‑assigned grading was a substantive component.

\subsection{Literature-Based Analysis}
Our literature-based analysis followed the PRISMA framework \cite{PAGE2021790} to ensure a clear and rigorous review process. We first defined the selection criteria based on the focus of our study on peer grading. We then conducted a comprehensive search across scholarly databases using targeted keywords. After removing duplicate records, we screened the studies based on their titles and abstracts, followed by a full-text review of the remaining papers. The final set of studies was analyzed to understand students' perceptions of peer grading, identify mitigation strategies for its negative aspects, and suggest directions for building a more pedagogically meaningful peer grading system.

\subsubsection{Source of Information}
\label{source_of_information}
The data for the literature-based analysis were collected from several reputable academic databases chosen for their broad coverage of scholarly work across multiple disciplines. These databases included ACM Digital Library, Web of Science, ERIC, PsycINFO, IEEE Xplore, and PubMed. Each source was selected because it provided access to high-quality, peer-reviewed research and was relevant to the topic of this study.

\subsubsection{Search Strategy and Eligibility Criteria}

An initial search was conducted across the selected databases using the keywords shown in the box below, together with the predefined inclusion and exclusion criteria. The eligibility criteria for this review were defined to ensure both relevance and quality. We included publications indexed in at least one of the six databases listed in Section \ref{source_of_information}. The review considered peer-reviewed journal articles, conference proceedings, book chapters, dissertations, and technical reports related to peer grading. To capture a broad and diverse range of approaches in this area, we included studies published between 2003 and 2025. For a study to be included, it had to report survey- or questionnaire-based evidence in which students expressed their opinions or perspectives on peer grading. We excluded studies that did not address at least one of our research questions, as well as non-English publications. Duplicate studies, books, and retracted papers were also excluded.
\vspace{2pt}
\begin{tcolorbox}[
colback=black!12,
colframe=black!80,
boxrule=0.5pt,
arc=2mm,
left=2mm,
right=2mm,
top=1mm,
bottom=1mm,
title=Search Keywords of Literature,
]
\footnotesize
("peer grading" OR "peer marking" OR "peer assessment" OR "peer scoring")
AND ("student perception*" OR "student attitude*" OR "student experience*" OR "student view*") 
AND (survey* OR questionnaire* OR self-report*)
\end{tcolorbox}

\subsubsection{Study Selection}

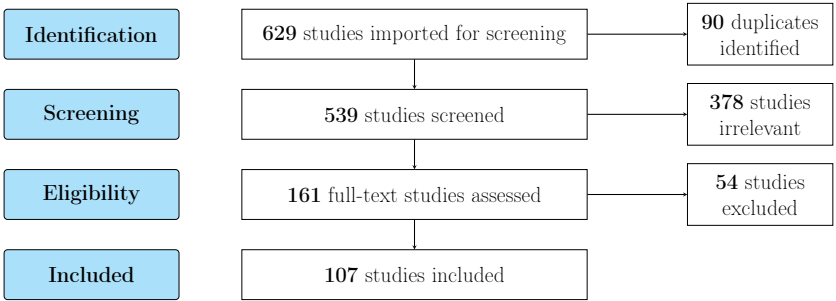
\begin{figure}[htbp]
    \centering
    \scalebox{0.33}{\tikzset{
    mynode/.style={
        draw,
        rectangle,
        align=center,
        text width=13cm,
        font=\Huge,
        inner sep=3ex,
        minimum height=2cm
    },
    mylabel/.style={
        draw,
        rectangle,
        align=center,
        rounded corners,
        font=\Huge\bfseries,
        inner sep=1ex,
        fill=cyan!30,
        minimum width=7cm,
        minimum height=2cm
    },
    arrow/.style={
        very thick,->,>=stealth
    }
}

\begin{tikzpicture}[
    scale=0.5,
    node distance=1.2cm,
    start chain=1 going below,
    every join/.style=arrow,
    ]

    % main vertical chain
    \node[mynode, on chain=1] (n1)
    {\textbf{629} studies imported for screening};

    \node[mynode, join, on chain=1] (n2)
    {\textbf{539} studies screened};

    \node[mynode, join, on chain=1] (n3)
    {\textbf{161} full-text studies assessed};

    \node[mynode, join, on chain=1] (n4)
    {\textbf{107} studies included};

    % right-side exclusion boxes
    \node[mynode, right=4cm of n1, text width=5cm] (r1)
    {\textbf{90} duplicates identified \\};
    \draw[arrow] (n1.east) -- (r1.west);

    \node[mynode, right=4cm of n2, text width=5cm] (r2)
    {\textbf{378} studies irrelevant};
    \draw[arrow] (n2.east) -- (r2.west);

    \node[mynode, right=4cm of n3, text width=5cm] (r3)
    {\textbf{54} studies excluded\\};
    \draw[arrow] (n3.east) -- (r3.west);

    % left-side blue labels aligned with each main box
    \node[mylabel, left=2.5cm of n1] (l1) {Identification};
    \node[mylabel, left=2.5cm of n2] (l2) {Screening};
    \node[mylabel, left=2.5cm of n3] (l3) {Eligibility};
    \node[mylabel, left=2.5cm of n4] (l4) {Included};

\end{tikzpicture}} % Scale as needed
    \caption{PRISMA flow diagram of the study selection process}
    \label{fig:prisma}
\end{figure}

 Our search returned 629 publications, of which 90 were identified as duplicates. After duplicate removal, 539 papers remained and were screened based on their abstracts to assess relevance. Following this stage, 161 studies were selected for full-text review. After the full-text review, 54 papers were excluded because they did not address our research questions. This resulted in a final set of 107 studies included in the review (Figure \ref{fig:prisma}).

\subsubsection{Quality Control:} Multiple authors were actively engaged throughout the research process to ensure the accuracy and trustworthiness of the studies included. Each paper was independently screened and reviewed by two authors. Any disagreement was followed by collaborative discussions to achieve agreement and reduce the likelihood of individual or methodological bias influencing the process. In the screening phase, Cohen's Kappa ($\kappa$) was 0.60, suggesting ``moderate'' agreement \cite{cohens:kappa, Landis:Koch:Kappa:Range}, while in the eligibility checking phase, the authors achieved substantial agreement ($\kappa$=0.68).

\subsection{Reddit Data}

For the Reddit data, we collected posts and comments from the Reddit archive using targeted keywords. A fine-tuned LLM was then used as an initial filtering step to remove irrelevant posts and comments. After validating the model’s performance on a subset of the data, the authors manually reviewed those LLM-classified relevant items and retained only the truly relevant ones for further analysis.

\subsubsection{Data Collection}
Reddit data were collected from a Reddit archive dump spanning from 2005 to 2025. We first downloaded the archive files and organized them by month. For each monthly dump, we searched for posts using the keywords ``peer grading'' and ``peer assessment.'' To broaden and diversify the dataset, we also collected comments from the retrieved posts, as students often share their opinions in the comments. This process yielded 659 posts and 6,607 comments.

\subsubsection{Annotation and Relevance Filtering:} From an initial manual inspection, we found that many of the retrieved posts and comments were irrelevant because the search was based only on keywords. Manually filtering the full dataset would have been too costly and time-consuming. To address this, we used an LLM to automate the relevance filtering process. For this purpose, we manually annotated a subset of the data and used it to fine-tune a Gemini 2.5 model.

The annotated dataset included 67 posts and 149 comments. Among them, 27 posts and 47 comments were labeled as relevant, while 40 posts and 102 comments were labeled as irrelevant. The items were not selected strictly in sequential or chronological order because an initial inspection indicated that doing so would produce a heavily imbalanced annotated dataset dominated by irrelevant instances. Instead, we used purposive sampling to include both likely relevant and irrelevant items, thereby reducing class imbalance while maintaining examples from both classes.

\subsubsection{Model Fine-Tuning and Relevance Filtering}
Using the annotated subset, we fine-tuned the Gemini 2.5 text classification model to identify likely relevant Reddit posts and comments at scale. After fine-tuning, we applied the model to the full dataset. Among 659 posts, the model labeled 203 as relevant. Similarly, among 6,607 comments, the model labeled 1,454 as relevant, while the remaining posts and comments were classified as irrelevant. To evaluate the effectiveness of the fine-tuned model, the authors independently annotated and validated a subset of 200 posts and comments. Table \ref{tab:confusion-matrix} presents the overall comparison between the human-annotated labels and the labels predicted by the model.

\begin{table}[htbp]
\centering
\caption{Confusion Matrix for Relevance Classification}
\label{tab:confusion-matrix}
\begin{tabular}{lcc}
\toprule
 & \multicolumn{2}{c}{\textbf{Predicted}} \\
\cmidrule(lr){2-3}
\textbf{Actual} & \textbf{Relevant} & \textbf{Irrelevant} \\
\midrule
\textbf{Relevant} & 66 & 1 \\
\textbf{Irrelevant} & 34 & 99 \\
\bottomrule
\end{tabular}
\end{table}

Table \ref{tab:confusion-matrix} presents the results of the relevance classification model. The model achieved an overall accuracy of 82.5\%. It performed very well in identifying relevant instances, achieving a recall of 98.5\%, which means that only a very small number of relevant posts and comments were missed. For the irrelevant class, the model achieved a recall of 74.4\% and a precision of 99.0\%. This indicates that when the model labeled a post or comment as irrelevant, the prediction was almost always correct. However, the precision for the relevant class was 66.0\%, showing that a considerable number of items predicted as relevant were actually irrelevant. Overall, these results suggest that the model was effective for filtering out irrelevant content while retaining most relevant cases, but manual verification of relevant items was still necessary.

\subsubsection{Final Dataset Construction}
Based on the spot-check results, we discarded the items classified as irrelevant by the fine-tuned model. For the items labeled as relevant, the authors manually inspected each instance and retained only those that were truly relevant. This process produced a final dataset of 114 relevant posts and 300 relevant comments. Inter-rater agreement for this manual review was $\kappa = 0.77$ for the Reddit posts and $\kappa = 0.95$ for the comments. The overall data processing procedure is summarized in Table \ref{tab:reddit-filtering}.

\begin{table}[htbp]
\centering
\caption{Reddit data reduction across the filtering process}
\label{tab:reddit-filtering}
\begin{tabular}{lcc}
\toprule
\textbf{Stage} & \textbf{Posts} & \textbf{Comments} \\
\midrule
Initial keyword-based retrieval     & 659 & 6,607 \\
After LLM-based filtering           & 203 & 1,454 \\
Final dataset after human validation & 114 & 300 \\
\bottomrule
\end{tabular}
\end{table}

\subsection{Encoding Student Perceptions and Mitigation Strategies}
To systematically organize students' perceptions of peer grading, we followed a bottom-up approach \cite{glaser1967discovery, nickerson2013method}. The authors independently analyzed all selected papers and the Reddit data. Each rater carefully examined how students described their perceptions of peer grading in both sources and independently identified positive and negative perceptions.

For mitigation strategies, we focused only on the papers so that the resulting strategies would be more grounded in prior research. The identified features were then refined, merged, or subdivided through iterative discussion, resulting in a consolidated set. Finally, the agreed-upon features were organized into 6 positive aspects, 7 negative aspects, and 8 mitigation strategies. Any disagreements that arose during this process were resolved by the last author, whose decision was treated as final.

For reporting, each literature paper and each Reddit item (post or comment) was coded based on whether it mentioned a given perception theme. A single paper or Reddit item could receive multiple codes when it expressed multiple perceptions. Therefore, the frequencies in the tables show how many literature papers or Reddit items mentioned each theme, and the summed frequencies across themes exceed the total number of papers or Reddit items. During theme encoding, mean inter-rater agreement between two raters across all themes was $\kappa = 0.67$ for the literature, $\kappa = 0.93$ for Reddit posts, and $\kappa = 0.88$ for Reddit comments. 

\subsection{Privacy and Ethical Considerations}
All Reddit analysis used publicly available discussion data and focused on aggregated thematic interpretation rather than user-level profiling. In reporting, comments were paraphrased or minimally quoted to help protect user privacy and reduce traceability. We did not attempt to identify or contact any Reddit users. The analysis focused on recurring themes in students' perceptions of peer grading rather than on individual users.

\section{Results}
This section presents a comparative analysis of students' perceptions of peer grading and the associated mitigation strategies, structured around our RQs.

\subsection{RQ1: Positive Perceptions of Peer Grading}
To answer RQ1, we identified six recurring patterns in students’ positive perceptions of peer grading across the literature and Reddit data, summarized below:
\vspace{1pt}

    \textbf{P1. Learning and understanding benefits:} Students often viewed peer grading as a useful way to improve their understanding of course content, assessment criteria, and performance expectations \cite{ Deslandes2021, Seneviratne2025}. Machanick \cite{Machanick2005} noted that peer grading helped students recognize their own mistakes. Similarly, Berrezueta et al. \cite{Berrezueta–Guzman2025} reported that reviewing peers’ work helped students explore different design ideas and learn from others’ solutions. Ayachi \cite{Ayachi2017} reported similar benefits for composition writing improvement. Deslandes and Hughes \cite{Deslandes2021} reported that 84.3\% of student markers agreed that acting as a peer marker helped them understand how the staff would mark their artifacts. Similar findings were reported in several other studies, including those by Jones \cite{Jones2010} and Kingsley \cite{Kingsley2010}. In addition to the benefits of reviewing, receiving feedback from peers also supported learning. For example, Verkade and Bryson-Richardson \cite{Verkade2013} found that peer feedback helped students improve their presentations. On Reddit, students similarly suggested that reviewing peers' work and receiving reviews were beneficial.
    
    \vspace{-2pt}
    \begin{tcolorbox}[redditquotebox]
    \footnotesize
    ``I’m a fan of peer grading in machine learning course because it let me see how other students approach the same problem and reinforces my understanding when I evaluate their solutions.'' \textbar\textbar\ ``I like it when someone points something out, because I can learn from their perspective.''
    \end{tcolorbox}
    
    \textbf{P2. Skill development benefits:} Students also perceived peer grading as beneficial for developing broader academic and professional skills. These included critical thinking, communication, reviewing ability, presentation, writing, and evaluative skills that extend beyond a single course. Lladó et al. \cite{Lladó2014} and Joh and Plakans \cite{Joh2025} reported that students developed critical thinking and evaluative skills through participation in peer grading tasks. Some studies also showed that peer grading helped students improve their presentation skills through peer feedback \cite{Peng2010, Gudiño2024}. In the Reddit discussions, one student mentioned that peer grading helped them learn how to critique others’ work.
    \vspace{-2pt}
    \begin{tcolorbox}[redditquotebox]
    \footnotesize
    ``I guess the one good thing that comes out of this is you get the perspective of other people that are not your Prof or TAs, and it also helps people learn how to critique other people's writing.''
    \end{tcolorbox}

   \textbf{P3. Engagement, motivation, and participation:} Peer grading was often seen as increasing students’ involvement in the learning process. It encouraged them to participate more actively, pay closer attention to the work, and feel more motivated and engaged in the course. For example, Callahan \cite{Callahan2008} reported that students perceived peer assessment as increasing their interest, participation, confidence, and desire to learn science. Ghosh and Skipper \cite{Ghosh2024} reported that students became more involved in learning because they actively participated in the assessment process. One Reddit comment suggested that peer grading increased engagement and participation, as students appreciated peer-graded exercises in the course.

    \begin{tcolorbox}[redditquotebox]
    \footnotesize
    ``I appreciate Coursera more for the peer-graded exercises, peer grading, and quizzes. Since some Google Cloud classes are available elsewhere without these exercises or certificates, I find them more valuable on Coursera.''
    \end{tcolorbox}
 
    \textbf{P4. Timeliness and turnaround benefits:} Another commonly reported benefit of peer grading was the speed of the process. Storjohann et al. \cite{Storjohann2019} reported that, among students, timely feedback on assignment performance was the second most frequently mentioned positive aspect, after faculty review of the rubric. Similarly, Callahan \cite{Callahan2008} reported that students perceived peer assessment as providing immediate feedback and a quick measure of their progress.

   \textbf{P5. Fairness, trust, and validity:} Some students expressed confidence in peer grading as a fair and credible evaluation method, although this often depended on how the process was designed. For example, Carvalho \cite{Carvalho2013} reported that although a substantial minority of students perceived peer assessment results as unfair, the majority of students (58.18\%) perceived the peer-assessment results as fair. Kumar et al. \cite{Kumar2019} found that students perceived anonymous online peer assessment as fair and valid, suggesting that anonymity can support students’ trust in peer grading. At the same time, the Reddit data suggest that students also viewed peer grading as generally fair and trustworthy in practice.
    
    \begin{tcolorbox}[redditquotebox]
    \footnotesize
    ``If you keep getting low grades from your peers, you should not just blame peer assessment but also reflect on your own work.'' \textbar\textbar\ ``If your work is good, you will probably receive good grades in most assignments, since it is graded by multiple people rather than just one.''
    \end{tcolorbox}

\textbf{P6. Social and collaborative benefits:} Students also valued peer grading for its social dimension. Many studies reported that it promoted interaction, discussion, collaboration, and shared learning, helping students feel more connected to their peers and the broader learning community \cite {Li2008, Käpylä2025}. Callahan \cite{Callahan2008} noted that peer grading encouraged collaboration, a sense of responsibility, and greater engagement with classmates. Similar reflections also appeared in the Reddit discussions, where students described peer review as a more relaxed and socially engaging experience and appreciated the opportunity to see others' work.
    
    \begin{tcolorbox}[redditquotebox]
    \footnotesize
    ``I kind of enjoyed the peer review. Less stressful, and you can leave your peers little notes like `good luck on the midterm'.'' \textbar\textbar\ ``In project-based classes like Ed Tech, it’s fascinating to see what others are doing and how.''\end{tcolorbox}

\begin{table}[htbp]
\centering
\caption{Distribution of positive aspects across literature and Reddit data. Values are shown as $n$ (\%). Percentages are calculated against the number of literature papers ($n$=107) and Reddit items ($n$=414).}
\label{tab:positive-aspects-distribution}
\footnotesize
\setlength{\tabcolsep}{4pt}
\resizebox{\linewidth}{!}{%
\begin{tabular}{lcccccc}
\toprule
\textbf{Source} & \textbf{P1} & \textbf{P2} & \textbf{P3} & \textbf{P4} & \textbf{P5} & \textbf{P6} \\
\midrule
Literature & 89 (83.2) & 85 (79.4) & 74 (69.2) & 10 (9.3) & 35 (32.7) & 62 (57.9) \\
Reddit Items & 22 (5.3) & 6 (1.4) & 2 (0.5) & 0 (0.0) & 63 (15.2) & 6 (1.4) \\
\bottomrule
\end{tabular}
}
\end{table}

Table \ref{tab:positive-aspects-distribution} shows that positive perceptions were distributed differently across the literature and Reddit data. In the literature, the most frequently reported benefits were learning and understanding benefits (P1, 83.2\%), skill development benefits (P2, 79.4\%), engagement and motivation (P3, 69.2\%), and social and collaborative benefits (P6, 57.9\%). In contrast, Reddit items mentioned fairness, trust, and validity (P5, 15.2\%) more often than the other positive themes, followed by learning and understanding benefits (P1, 5.3\%). Timeliness and turnaround benefits (P4) did not appear in the Reddit data. Overall, the literature emphasized the learning and collaborative value of peer grading more strongly, whereas Reddit discussions contained relatively fewer positive themes and focused more on fairness and validity.

\subsection{RQ2: Negative Perceptions of Peer Grading}

For RQ2, we identified seven common negative perception patterns across the literature and Reddit data, summarized below:

\textbf{N1. Accuracy and reliability concerns:} One of the most common concerns students expressed about peer grading was related to accuracy and reliability. Students often reported that peer-assigned grades were inaccurate, inconsistent across reviewers, sometimes arbitrary, and less reliable than instructor grading. Several studies also noted tendencies toward over-marking or under-marking, as well as concerns that peer evaluations did not reflect staff-level judgment. For example, Verkade and Bryson-Richardson \cite{Verkade2013} reported that peer grading was limited by grade inflation, reluctance to assign low marks, and inconsistent evaluation. Hammer et al. \cite{Hammer2010} also reported a lack of grading uniformity, while Kingsley \cite{Kingsley2010} found that peer grading suffered from low reliability, leniency bias, and lack of expertise. Toll and Wingkvist \cite{Toll2017} likewise noted that the strongest concerns were incorrect feedback and incorrect scores from peers, with 27\% of students reporting incorrect feedback and 31\% reporting incorrect scores. On Reddit, several students echoed the same problem.

\begin{tcolorbox}[redditquotebox]
\footnotesize
``I got my written assignment back: one grader gave 90/90, another 17/90, and another 72/90. Did they even read the same paper?'' \textbar\textbar\ ``I honestly think some people do not even read the assignment and just grade it randomly.''
\end{tcolorbox}

A more recent concern about accuracy raised by students was the use of automated tools, particularly ChatGPT, in the grading process. While such tools may introduce some level of consistency, students perceived them as unreliable or inappropriate for evaluating peer work.

\begin{tcolorbox}[redditquotebox]
\footnotesize
``Now peers run essays through ChatGPT, which gives a solid 9 or 10 essay a 7 just because it does not cover everything that could fit in 5000 words, even though the limit was only 500.'' \textbar\textbar\ ``Two peers graded my written assignment and both gave it 6/10. I have never seen such consistent grading. One of the comments even started with `The author'.''
\end{tcolorbox}

\textbf{N2. Fairness and impartiality concerns:} Students also expressed concerns about the fairness of peer grading frequently. Students reported issues such as unfair grading driven by personal bias, friendship-based marking, bias stemming from personal relationships or conflicts, and intentional low scoring driven by competitiveness. Carvalho \cite{Carvalho2013} found that about 30\% of students belonged to a negative-experience group that viewed peer grades as unfair, often because of friendship marking and conflict. These concerns have also been echoed in several other studies \cite{ Verkade2013, Kingsley2010}. In Reddit discussions, students also described unfair grading behaviors such as retaliation, collusion, and strategic grading.

\begin{tcolorbox}[redditquotebox]
\footnotesize
``I got three peer grades: a 5, a 4, and a 1. The 1 came from someone I had been arguing with'' \textbar\textbar\ ``If the grades I give others affect my own GPA, why should I not just give everyone 10/10? Every little thing counts, and it feels fair game.''
\end{tcolorbox}

\textbf{N3. Insufficient assessor preparedness:} Another major concern was the limited preparedness and expertise of peer assessors. Students often felt that their peers lacked the subject knowledge, experience, confidence, and rubric understanding needed to provide accurate and consistent evaluations. They also reported difficulty judging certain criteria and assigning appropriate numerical grades. Bauer et al. \cite{Bauer2009} reported that students found some review criteria, particularly sources and content, difficult to assess. Seneviratne et al. \cite{Seneviratne2025} found that 34.88\% of student assessors reported that they did not feel they had the skills and knowledge needed to assess their peers. Other studies also reported low confidence in both students’ own and their peers’ ability to assess work accurately \cite{Vo2023}.
Similar concerns were also reflected in Reddit discussions, where students questioned the competence and preparedness of peer graders. Many described peers as unqualified, lacking understanding of the rubric, or unable to provide meaningful evaluation.

\begin{tcolorbox}[redditquotebox]
\footnotesize
``Because who would want their grades decided by a random group of peers with the expertise of a potato?'' \textbar\textbar\ ``Instead, it just seems to give unqualified, wannabe tyrants a platform to jeopardize other students’ standing.'' \textbar\textbar\ ``I think the university should intervene more, since unqualified individuals often are not even familiar with the grading rubric.''
\end{tcolorbox}

\textbf{N4. Poor feedback quality and weak learning benefits:} Unlike the previous three concerns, which were mainly tied to the summative aspect of peer grading, this concern centered on its formative value and learning benefits. Students often described peer feedback as unclear, minimal, vague, contradictory, or insufficient to support improvement. Bratkovich \cite{Bratkovich2014} found that student feedback was more surface-level than teacher feedback, focusing more on language issues than elaboration and idea development. Kobayashi \cite{Kobayashi2020} similarly found that overly nice and noncritical feedback reduced its usefulness. Other studies also reported incorrect feedback from peers \cite{Toll2017}, as well as feedback that was insufficiently constructive \cite{Storjohann2019}. Kay \cite{Kay2022} further reported that, although overall perceptions of peer grading were positive, 30.82\% of students disagreed that it supported their learning. In Reddit discussions, some students also mentioned the same issues.

\begin{tcolorbox}[redditquotebox]
\footnotesize
``Peer grades often feel harsh and give me little useful feedback beyond something like `see solutions'.'' \textbar\textbar\ ``I dislike peer-graded homework because the feedback is often unhelpful and sometimes wrong.'' \textbar\textbar\ ``I once got comments like `great job' but still received a bad grade. I am less upset about the marks than about not getting enough feedback to improve.''
\end{tcolorbox}

\textbf{N5. Low motivation and engagement:} Students sometimes reported low motivation toward peer grading tasks. Common concerns included limited interest in reading others’ work, spending little time on reviews, not taking the task seriously, and reducing effort because easy marks were expected regardless of review quality. Toll and Wingkvist \cite{Toll2017} identified time and effort as contributors to low motivation in peer review. Joh and Plakans \cite{Joh2025} reported that negative or harsh peer feedback can cause stress and lower motivation among some students. Similar patterns also appeared in Reddit discussions, where students described rushing through required reviews, assigning full marks just to finish quickly, and losing interest when peers seemed not to read submissions carefully.

\begin{tcolorbox}[redditquotebox]
\footnotesize
``Even though peer reviews are supposed to be taken seriously, I usually give everyone full marks because people already have enough going on, and it is not life or death.'' \textbar\textbar\ ``The number of people who were not even reading my papers was insane. They turned peer grading into a joke.''
\end{tcolorbox}

Reddit discussions also suggested that the visible misuse of AI in peer submissions could further demotivate students. When students noticed signs of AI-generated work, they sometimes felt less willing to engage seriously with the review process and preferred to avoid reviewing such submissions altogether.
\begin{tcolorbox}[redditquotebox]
\footnotesize
    ``I saw `Copy code Python' in one post and a chatbot prompt in another. If I have a choice, I do not grade such assignments and instead review posts written by people who actually did the work.''
\end{tcolorbox}

\textbf{N6. Emotional and interpersonal burden:} Students often reported emotional stress related to peer grading. Common concerns included fear of losing face in front of peers, anxiety about receiving incorrect or unfair evaluations, anxiety about assigning low grades to peers, discomfort being assessed by people they knew, and feeling discouraged or upset by harsh feedback. Deslandes and Hughes \cite{Deslandes2021} found that although most students valued peer marking, some felt uncomfortable when they knew the peer markers or when multiple markers were present. Hammer et al. \cite{Hammer2010} reported increased anxiety caused by competition and strategic grading. Seneviratne et al. \cite{Seneviratne2025} found that 38.36\% of student assessors were reluctant to give low marks, and 22.08\% reported difficulty separating personal feelings from peer evaluation. Kobayashi \cite{Kobayashi2020} similarly found that students were reluctant to criticize their peers honestly. Stancic et al. \cite{Stancic2021} further noted that students were concerned about the effect of peer grading on personal relationships. Similar concerns emerged in Reddit discussions, where students described the emotional and interpersonal burden of peer grading, including guilt, anxiety, embarrassment, and fear of social backlash.

\begin{tcolorbox}[redditquotebox]
\footnotesize
``Peer grading is not too bad, but I still feel bad when I give someone a lower grade than they deserve because of a brain fart.'' \textbar\textbar\ 
``As a socially anxious person, I would worry about getting nasty comments on my project.'' \textbar\textbar\ 
``I was terrible at math as a child, and it felt embarrassing and shameful to have my peers see that firsthand.'' \textbar\textbar\ 
``There is a burden in being seen as the negative one or constant complainer, even when the criticism is productive.''
\end{tcolorbox}

\textbf{N7. Workload, time, and coordination challenges:} Students also reported workload- and time-related challenges in the peer grading process. Common concerns included the high burden of reviewing multiple peers, limited time to provide thoughtful feedback, difficulty coordinating with others during the process, and the overall effort required to complete peer reviews. Siow et al. \cite{Siow2015} reported that 43\% of students found the peer grading process time-consuming. Similarly, Badea and Popescu \cite{Badea2024} found that students viewed peer grading as time-intensive and noted that using it across many subjects would create a burden. Toll and Wingkvist \cite{Toll2017} also reported that some students viewed peer grading as time-consuming, which may have contributed to lower motivation toward reviewing peers’ reports. Deslandes and Hughes \cite{Deslandes2021} further noted that students experienced time pressure when providing feedback, while some students in Käpylä and Palvalin’s study \cite{Käpylä2025} described giving feedback as laborious and difficult.

\begin{table}[htbp]
\centering
\caption{Distribution of negative aspects across literature and Reddit data. Values are $n$ (\%), with percentages calculated against literature papers ($n$=107) and Reddit items ($n$=414).}
\label{tab:negative-aspects-distribution}
\footnotesize
\setlength{\tabcolsep}{4pt}
\resizebox{\linewidth}{!}{%
\begin{tabular}{lccccccc}
\toprule
\textbf{Source} & \textbf{N1} & \textbf{N2} & \textbf{N3} & \textbf{N4} & \textbf{N5} & \textbf{N6} & \textbf{N7} \\
\midrule
Literature & 50 (46.7) & 35 (32.7) & 35 (32.7) & 31 (29.0) & 10 (9.3) & 45 (42.1) & 29 (27.1) \\
Reddit Items & 239 (57.7) & 94 (22.7) & 58 (14.0) & 71 (17.1) & 30 (7.2) & 19 (4.6) & 18 (4.3) \\
\bottomrule
\end{tabular}
}
\end{table}

Table \ref{tab:negative-aspects-distribution} shows that negative perceptions were prominent in both the literature and Reddit data, but with different distributions. In the literature, the most frequently reported concerns were accuracy and reliability (N1, 46.7\%), emotional and interpersonal burden (N6, 42.1\%), fairness and impartiality (N2, 32.7\%), and insufficient assessor preparedness (N3, 32.7\%). In contrast, Reddit items were most strongly associated with accuracy and reliability concerns (N1, 57.7\%), followed by fairness and impartiality concerns (N2, 22.7\%), poor feedback quality and weak learning benefits (N4, 17.1\%), and insufficient assessor preparedness (N3, 14.0\%). This pattern suggests that the literature reflects a broader spread of negative themes, whereas Reddit discussions are more concentrated on accuracy and reliability concerns.

\subsection{RQ3: Mitigation Strategies}

Although students reported several negative perceptions of peer grading, the literature also identified strategies to address them. In our analysis, we identified eight recurring mitigation strategies, described below.

\textbf{M1. Training and preparation:} Training and preparing students before peer grading can help address several common problems. This preparation may include practice grading with example papers or calibration tasks, repeated activities to build comfort, orientation on how to give good feedback and apply grading criteria, brief verbal explanations of the system, online tutorials on providing effective peer feedback, and explicit staff discussions about the value of feedback. Gudiño et al. \cite{Gudiño2024} reported greater confidence in peer grading and less relationship-based influence after preparation and practice. Similarly, Sadeghi and Khonbi \cite{Sadeghi2015} noted that training and practice can improve reliability and validity and foster more positive student attitudes toward peer grading.

 \textbf{M2. Rubrics and structured grading:} Using clear grading guides instead of open-ended scales can help reduce peer-grading problems. Helpful approaches include detailed rubrics with explicit criteria and performance levels, structured scoring guides such as checklists, and clear instructor-defined review criteria. Berrezueta-Guzman et al. \cite{Berrezueta–Guzman2025} described a rubric-based, anonymized peer-review process in which students often linked fairness to consistent rubric use. de Raadt et al. \cite{deRaadt2007} similarly argued that clear, objective, task-completion-focused criteria could reduce ambiguity and improve consistency among peer reviewers.
 
\textbf{M3. Anonymity and blind review:} Concerns about fairness, social discomfort, and anxiety can be reduced through anonymous peer grading. Depending on the context and purpose, anonymity may be implemented through either single-blind or double-blind review. Kobayashi \cite{Kobayashi2020} found that anonymity helped reduce students’ anxiety and fear of hurting peers’ feelings. Venables and Summit \cite{Venables2003} also used blind review and argued that anonymity was important for fairness in peer grading. Similarly, Vo and Nguyen \cite{Vo2023} suggested that anonymity can reduce friendship-based bias and improve validity and reliability, while Biton \cite{Biton2025} argued that it can encourage more honest and objective feedback.

\textbf{M4. Assessment by multiple reviewers:} Because peer grading can be affected by inaccuracy and bias, involving multiple reviewers and combining their scores can improve the accuracy of final grades. This can be done through simple averaging, filtering out noisy ratings, or more advanced aggregation methods\cite{paul2025}. Landry et al. \cite{Landry2015} noted that although peer grading may be inconsistent at the individual level, it can still support learning and produce reasonably accurate results when multiple reviewers are involved. Verkade and Bryson-Richardson \cite{Verkade2013} showed that averaging marks from many peers brought peer scores closer to academic marks, reducing individual rater noise and improving accuracy. 

\textbf{M5. Instructor oversight and moderation:} Instructor oversight and moderation can help address many peer-grading problems. Helpful approaches include identifying unfair or unreliable reviews, intervening in grading disputes, monitoring the review process, conducting discussion sessions, providing on-demand support, and combining peer and instructor evaluations. de Raadt et al. \cite{deRaadt2007} introduced instructor moderation when peer reviews conflicted or appeared inaccurate. It helped improve the validity and fairness of final marks. Storjohann et al. \cite{Storjohann2019} found that faculty-led review and reflection sessions supported fairer scoring and helped students grade their peers’ work more effectively. Toll and Wingkvist \cite{Toll2017} included teacher intervention within a tool-supported peer review process to manage disputes, and reported that students experienced fewer problems with incorrect feedback and scoring than they had expected.

\textbf{M6. Incentives and accountability:} Incentives and accountability can help reduce problems such as low motivation and poor-quality feedback. Useful strategies include rewarding participation, recognizing high-quality feedback, and giving marks for the quality of peer grading. Toll and Wingkvist \cite{Toll2017} suggested that grading reviews and allowing students to rate the feedback they received helped motivate better reviews. It also reduced concerns about incorrect feedback. Gillanders et al. \cite{Gillanders2020} used a loss-aversion incentive. Students could lose a small number of marks if they did not assess a peer’s work accurately. This increased engagement and supported critical thinking during peer grading.

\textbf{M7. Time and workload management:} Good time and workload management can help reduce several peer-grading problems, including poor feedback quality and workload-related stress during the process. Helpful strategies include giving students enough time to review, limiting the number of assigned reviews, and reducing workload through careful design of assessment criteria. Wang et al. \cite{Wang2020} reported that time constraints shaped students’ attitudes toward online peer grading and recommended allowing more time for review. Peng \cite{Peng2010} similarly argued that using fewer assessment criteria can make peer grading more manageable for students by reducing confusion and time demands.

\textbf{M8. Technology and tool support:} Technology and tool support can help address some peer-grading problems and improve the overall experience. Lai and Lan \cite{Lai2006} showed that a computer-supported agent-negotiation system reduced personal bias, subjective judgment, and unfair grading. It also improved flexibility and equity in the process. Toll and Wingkvist \cite{Toll2017} similarly reported that, within a tool-supported peer-review process, students experienced fewer problems with incorrect scores and feedback than they had expected, and their overall attitudes toward peer review improved. Sullivan and Watson \cite{Sullivan2015} further found that LMS-based peer grading made the process more practical and easier to manage in hybrid and online courses. Several studies also highlighted the role of technology in speeding up feedback and grade return. For example, Chen \cite{Chen2010} found that the mobile peer-assessment system improved portability, time efficiency, and access to immediate feedback, while Li et al. \cite{Li2008} reported that their web-based systems enabled more timely feedback for students.

\begin{table}[t]
\centering
\small
\setlength{\tabcolsep}{6pt}
\caption{\textbf{Negative perceptions and mitigation strategies.} A checkmark indicates that the strategy was suggested in at least one study for that perception.}
\label{tab:negative_mitigation_map}
\begin{tabular}{lcccccccc}
\toprule
 & \textbf{M1} & \textbf{M2} & \textbf{M3} & \textbf{M4} & \textbf{M5} & \textbf{M6} & \textbf{M7} & \textbf{M8} \\
\midrule
\textbf{N1} & $\checkmark$ & $\checkmark$ & $\checkmark$ & $\checkmark$ & $\checkmark$ & - & - & $\checkmark$ \\
\textbf{N2} & $\checkmark$ & $\checkmark$ & $\checkmark$ & $\checkmark$ & $\checkmark$ & - & - & $\checkmark$ \\
\textbf{N3} & $\checkmark$ & $\checkmark$ & - & - & $\checkmark$ & - & - & - \\
\textbf{N4} & $\checkmark$ & $\checkmark$ & - & $\checkmark$ & $\checkmark$ & $\checkmark$ & $\checkmark$ & $\checkmark$ \\
\textbf{N5} & $\checkmark$ & - & - & - & $\checkmark$ & $\checkmark$ & - & - \\
\textbf{N6} & $\checkmark$ & $\checkmark$ & $\checkmark$ & $\checkmark$ & $\checkmark$ & - & - & - \\
\textbf{N7} & $\checkmark$ & - & - & - & $\checkmark$ & - & $\checkmark$ & $\checkmark$ \\
\bottomrule
\end{tabular}
\end{table}

Table~\ref{tab:negative_mitigation_map} shows that M1 (training and preparation) and M5 (instructor oversight and moderation) are the most central strategies, as they address all seven negative perceptions. This suggests that students are more likely to accept peer grading when they are properly prepared and when instructors remain involved in monitoring and moderating the process. The table also shows that N1 (accuracy and reliability concerns), N2 (fairness and impartiality concerns), N4 (poor feedback quality and weak learning benefits), and N6 (emotional and interpersonal burden) are linked to a wide range of strategies, suggesting that these concerns are more complex and often require multiple forms of support. By contrast, N3 (insufficient assessor preparedness) and N5 (low motivation and engagement) are associated with only a few strategies. Among the more targeted strategies, M2 (rubrics, criteria, and structured grading) helps reduce ambiguity, M3 (anonymity and blind review) helps reduce social pressure and bias, M4 (assessment by multiple reviewers) improves robustness, M6 (incentives and accountability) encourages engagement, M7 (time and workload management) makes the process more manageable and M8 (technology and tool support) addresses concerns related to accuracy, fairness, feedback quality, and workload.

\section{Discussion}

Our findings show that students perceive peer grading as both beneficial and problematic. They often associate it with learning, skill development, collaboration, engagement, and sometimes faster feedback. At the same time, they repeatedly raised concerns about accuracy, fairness, assessor preparedness, feedback quality, emotional burden, and workload. Overall, students seem to judge peer grading less as an abstract idea and more by whether its implementation feels useful, fair, and trustworthy. Trust appears central to student acceptance. Although peer grading can support learning and collaboration, many students seem to evaluate it first through fairness, validity, and reliability. When trust is weak, its assessment role appears to overshadow its learning value.

A notable pattern is the contrast between the literature and Reddit discussions. The literature presents a more balanced picture, with both positive and negative perceptions appearing regularly, whereas Reddit discussions were much more dominated by negative experiences. This likely reflects the difference between structured classroom settings reported in research and informal online spaces where students are likely to voice frustration. Although Reddit does not fully represent classroom reality, it still highlights the problems students find most memorable and worth expressing.

The mitigation analysis shows that many common concerns are avoidable. They can often be reduced through careful design choices such as training, rubrics, anonymity, multiple reviewers, instructor moderation, incentives, workload management, and technology support. Among these, M5 (instructor oversight and moderation) and  M1 (training and preparation) appear to be the most central because they address all seven major negative perceptions. This shows that the success of peer grading depends less on peer involvement and more on how well the process is designed and supervised.

The findings also point to an emerging concern related to AI use in peer grading. Reddit discussions suggest that some students suspect peers of relying on tools such as ChatGPT to evaluate work or generate feedback. Rather than improving the process, this appeared to deepen concerns about inaccuracy, generic comments, and lack of genuine engagement. This suggests that while technology may support coordination and efficiency, it may also create new trust issues when students feel that human judgment is being replaced.

The literature review was limited to English-language studies published between 2003 and 2025, while the Reddit analysis relied on keyword-based retrieval from public discussions. As a result, some relevant perspectives may have been missed. In addition, perceptions were drawn from both formal studies and informal online discussions, which may differ in context and style. Moreover, Reddit offers only a complementary, not fully representative, view of classroom reality.

\section{Conclusion}

This study provides a mixed-source view of students’ perceptions of peer grading by combining evidence from literature and Reddit discussions. The findings show that peer grading can support learning, skill development, and collaboration, but students also report important concerns about fairness, reliability, feedback quality, and assessor preparedness. Overall, the results suggest that peer grading is most positively received when students are adequately trained, the process is clearly structured, and instructors remain actively involved.

\bibliographystyle{splncs04}
\bibliography{references}

\end{document}